\documentclass[12pt]{iopart}
\usepackage{iopams,graphicx,bm}  
\newcommand\DTO{$\rm Dy_2Ti_2O_7$}
\newcommand\ETO{$\rm Er_2Ti_2O_7$}

\newcommand\GTO{$\rm Gd_2Ti_2O_7$}
\newcommand\HTO{$\rm Ho_2Ti_2O_7$}
\newcommand\pyro{pyrochlore}

\newcommand\PSJ[3]{{#3} {\it J. Phys. Soc. Japan} {\bf {#1}} {#2}}
\newcommand\PRB[3]{{#3} {\it Phys. Rev. B} {\bf {#1}} {#2}}
\newcommand\PCM[3]{{#3} {\it J. Phys.: Cond. Matt.} {\bf {#1}} {#2}}

\newcommand\PPRL[3]{{#3} {\it Phys. Rev. Lett.} {\bf {#1}} {#2}}
\newcommand\MH{Magnetisation}
\newcommand\mH{magnetisation}
\begin{document}
\title[Titanium pyrochlore magnets: magnetisation measurements]{Titanium pyrochlore magnets: how much can be learned from magnetisation measurements?}
\author{O A Petrenko, M R Lees and G Balakrishnan}
\address{University of Warwick, Department of Physics, Coventry, CV4~7AL, UK}
\ead{o.petrenko@warwick.ac.uk}
\begin{abstract}
We report magnetisation data for several titanium \pyro\ systems measured down to 0.5~K.
The measurements, performed on single crystal samples in fields of up to 7 tesla, have captured the essential elements of the previously reported phase transitions in these compounds and have also revealed additional important features overlooked previously either because of the insufficiently low temperatures used, or due to limitations imposed by polycrystalline samples.
For the spin-ice \pyro s \DTO\ and \HTO, an unusually slow relaxation of the \mH\ has been observed in lower fields, while the \mH\ process in higher fields is essentially hysteresis-free and does not depend on sample history.
For the  $XY$ \pyro\ \ETO, the magnetic susceptibility shows nearly-diverging behaviour on approach to a critical field, $H_C=13.5$~kOe, above which the \mH\ does not saturate but continues to grow at a significant rate.
For the Heisenberg \pyro\ \GTO, the magnetic susceptibility shows a pronounced change of slope at both transition temperatures,  $T_{N1}=1.02$~K and $T_{N2}=0.74$~K, contrary to the earlier reports.
\end{abstract}

\section{Introduction}
Since the publication 10 years ago of a paper on the bulk \mH\ of the heavy rare earth titanate \pyro s~\cite{Bramwell_JPCM_2000}, significant progress has been made in the understanding of the magnetic properties of the various members of this family~\cite{Gardner_review_2010}.
This progress is due to the utilisation of both advanced experimental techniques (such as, for example, diffuse neutron scattering with polarisation analysis~\cite{Fennell_Science_2009}) and modern numerical and theoretical methods.
Despite often being perceived as simple and straightforward, \mH\ measurements capture nicely the physics of titanium \pyro s. 
We revisit and extend the bulk \mH\ studies of the spin-ice systems \DTO\ and \HTO, an $XY$ antiferromagnet \ETO, and a Heisenberg \pyro\ \GTO, by performing the measurements (\textit{i}) on high quality single crystals  (\textit{ii}) down to 0.5~K.
The \mH\ values, the rate of their change with applied field, and especially the anisotropy observed, all give real insights into the nature of the magnetic ground states and the field-induced transitions in the titanium \pyro s.
Particular attention is paid to sample history dependence, as well as to the time dependence of the \mH, as unusually slow relaxation processes are discovered at low temperatures.

Given the nature of this special issue of {\it Journal of Physics: Condensed Matter} on geometrically frustrated magnetism an extensive general introduction to the \pyro\ systems is omitted.
Instead, after a brief description of the experimental details, we proceed to the presentation of the results and compare them to previously published data where necessary.

\section{Experimental procedures}
Single crystals of the \pyro\ magnets were grown by the floating zone technique, using an infrared image furnace~\cite{Balakrishnan_JPCM_1998}.
Small thin plates of a characteristic size $2\times 2\times 0.5$~mm$^3$ (typically containing the (110)-plane) were cut from the original larger samples.
The plates were mounted in such a way that the magnetic field was applied in the sample plane in order to minimize de\mH\ effects~\cite{demag}.
The principal axes of the samples were determined using the X-ray diffraction Laue technique; the crystals were aligned to within an accuracy of 2$^{\circ}$.
In this article we report the \mH\ data for a field applied along the $[111]$ axis, which is an ``easy-axis'' direction for the spin-ice systems  \HTO\ and \DTO\ and a ``hard-axis" direction for the $XY$ antiferromagnet  \ETO.
 
\MH\ measurements were made down to 0.5~K in applied magnetic fields of up to 7~T using a Quantum Design Magnetic Properties Measurement System SQuID magnetometer along with an i-Quantum $^3$He insert.
The \mH\ was measured both as a function of temperature in a constant magnetic field and as a function of applied field at constant temperature. 

\section{Experimental results and discussion}
\subsection{Spin-ice compounds \HTO\ and \DTO}
\begin{figure}
\includegraphics[width=\columnwidth]{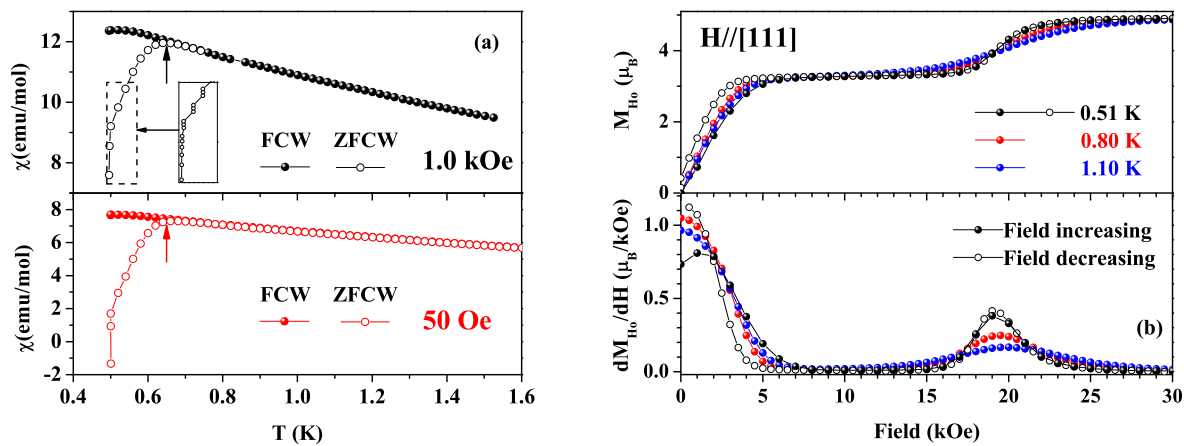}
\caption{(a) Temperature dependence of the magnetic susceptibility of \HTO\ single crystal measured on warming for field cooled (solid symbols) and zero field cooled (open symbols) samples.
		Each point is obtained by averaging three consecutive measurements, each of them typically taking 1-2 minutes.
		The inset emphasises a pronounced time dependence of the ZFC susceptibility for $T_C<0.65$~K (marked by arrows) by showing raw data in a selected region.
		(b) Field dependence of the \mH\ measured for increasing and decreasing magnetic field (shown by solid and open symbols respectively).
		The $dM/dH$ curves obtained from the magnetisation curves at different temperatures show that temperature effects are most pronounced at both ends of a plateau.}
\label{Fig1} 
\end{figure}
The temperature and field dependence of the \mH\ of \HTO\ is shown in Fig.~\ref{Fig1}.
Below a temperature of approximately 0.65~K the magnetic susceptibility of \HTO\ is very sensitive to the sample history, as the $\chi(T)$ curves for field cooled (FC) and zero field cooled (ZFC) samples differ significantly, particularly for lower applied fields.
Below $T_C=0.65$~K, the ZFC susceptibility shows a pronounced time dependence -- it continues to grow with time even when the temperature remains constant.
Moreover, if the applied field is as weak as 50~Oe, the susceptibility measured at $T=0.5$~K is initially negative.
It grows rapidly with temperature and time and joins the FC susceptibility at $T_C$ (see Fig.~\ref{Fig1}a).
A natural explanation of the observed phenomenon is linked to the presence of a small, typically less than 5~Oe, {\it frozen} field always present in a cryomagnet.
Therefore a ZFC protocol is in fact a cooling in a very weak field of opposite polarity to the main applied field.
It is still surprising to observe, at temperatures near 0.5~K, a process of \mH\ reversal on a time scale of several minutes.

\begin{figure}
\includegraphics[width=0.5\columnwidth]{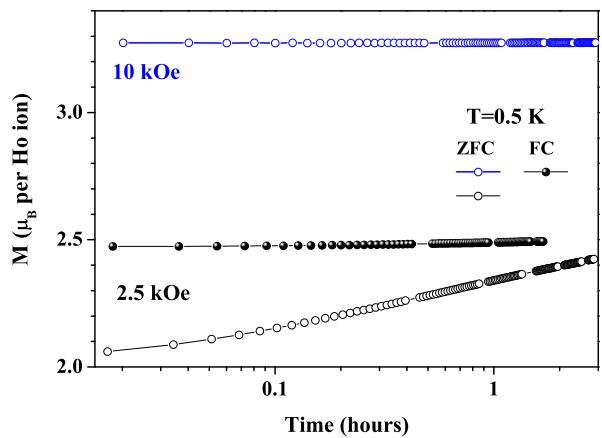}
\caption{Time dependence of the \mH\ of \HTO\ single crystals for a field applied along the $[111]$ direction at $T=0.5$~K.
		FC and ZFC measurements are shown by solid and open symbols respectively.
		The change in the ZFC signal at 2.5~kOe over a period of nearly three hours is 20\%, while the changes in the FC signal at 2.5~kOe and the ZFC signal at 10~kOe are 0.8\% and 0.07\% respectively.}
\label{Fig2} 
\end{figure}

For \HTO, we have previously reported \mH\ curves for all three principal directions of an applied magnetic field at temperatures down to 1.6~K~\cite{Petrenko_PRB_2003}.
Lowering the temperature to 0.5~K makes the plateau for $H\parallel [111]$ much more pronounced (see Fig.~\ref{Fig1}b).
Comparing the results obtained for increasing and decreasing applied fields, one has to conclude that the \mH\ process in higher fields is essentially hysteresis-free, while a significant hysteresis is evident in a field of less than 5~kOe.
As can been seen in Fig.~\ref{Fig2}, the \mH\ has a pronounced time dependence in weaker magnetic fields.
For example, if the field applied to a ZFC sample is rapidly raised to 2.5~kOe and stabilised, the first measurement immediately after field stabilisation  gives 2.02~$\mu_B$ per Ho ion, while measurements 30 and 120 minutes later give 2.28 and 2.40~$\mu_B$ per Ho ion respectively.
Technical limitations do not allow us to maintain a constant sample temperature ($T<0.51$~K) for more than 3 hours.
However, it is very likely that after several hours, the measurements would return a value very close to the FC \mH\ of 2.50~$\mu_B$ per Ho ion.
The \mH\ shows no significant time dependence in fields of more than 5~kOe or at temperatures above 0.65~K.

\begin{figure}
\includegraphics[width=0.5\columnwidth]{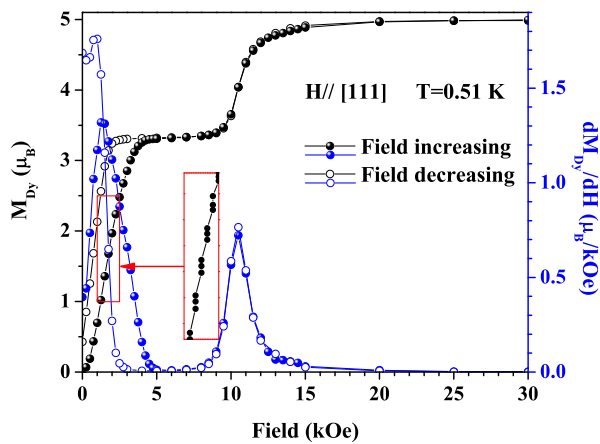}
\caption{Field dependence of the \mH\ $M$ as well as of the $dM/dH$ curves for \DTO.
		Each point is obtained by averaging three consecutive measurements.
		The inset emphasises a pronounced time dependence of the \mH\ for lower fields by showing raw data in a selected region.
		}
\label{Fig3} 
\end{figure}

The low-temperature \mH\ process in the spin-ice compound \DTO\ has been extensively studied previously.
For $H\parallel [111]$ there is an agreement between our data shown in Fig.~\ref{Fig3} and the data originally reported by Sakakibara's group~\cite{Sakakibara_group}.
In common with what is observed for \HTO, the \mH\ process in \DTO\ also consists of four distinct regions: (\textit{i}) a sharp rise in fields below 5~kOe; (\textit{ii}) a plateau (often labelled as the Kagome ice state) in intermediate fields, (\textit{iii}) a further sharp rise in a field of 10.5~kOe for \DTO\ and 19.0~kOe for \HTO; (\textit{iv}) another plateau in higher fields.
The second plateau, which is sometimes referred to as the saturation \mH, is associated with a state where the spin-ice rules are broken by the force of the applied field.
Hysteretic behaviour and pronounced time dependence of \mH\ at lower temperatures are only seen in the first low field region (see inset in Fig.~\ref{Fig3}).
The time and history dependence is significant only at $T<T_C\approx 0.65$~K, the same temperature as has been reported previously as a bifurcation point between the FC and ZFC powder \mH\ data~\cite{Snyder_PRB_2004}.
According to the latest reports~\cite{Slobinsky_arXIv_2010}, on lowering the temperature to $\sim 0.1-0.2$~K, pronounced non-equilibrium processes are observed, culminating in the appearance of sharp \mH\ steps accompanied by similarly sharp peaks in the temperature of the sample.
The low-temperature behaviour is ascribed to the field-driven motion of the magnetic monopole excitations~\cite{Slobinsky_arXIv_2010}.
In would be interesting to check if the observed time-dependence at intermediate-temperatures in both \DTO\ and \HTO\ could also be described in terms of monopole movement~\cite{monopole}. 

\subsection{$XY$ antiferromagnet \ETO}
\begin{figure}
\includegraphics[width=\columnwidth]{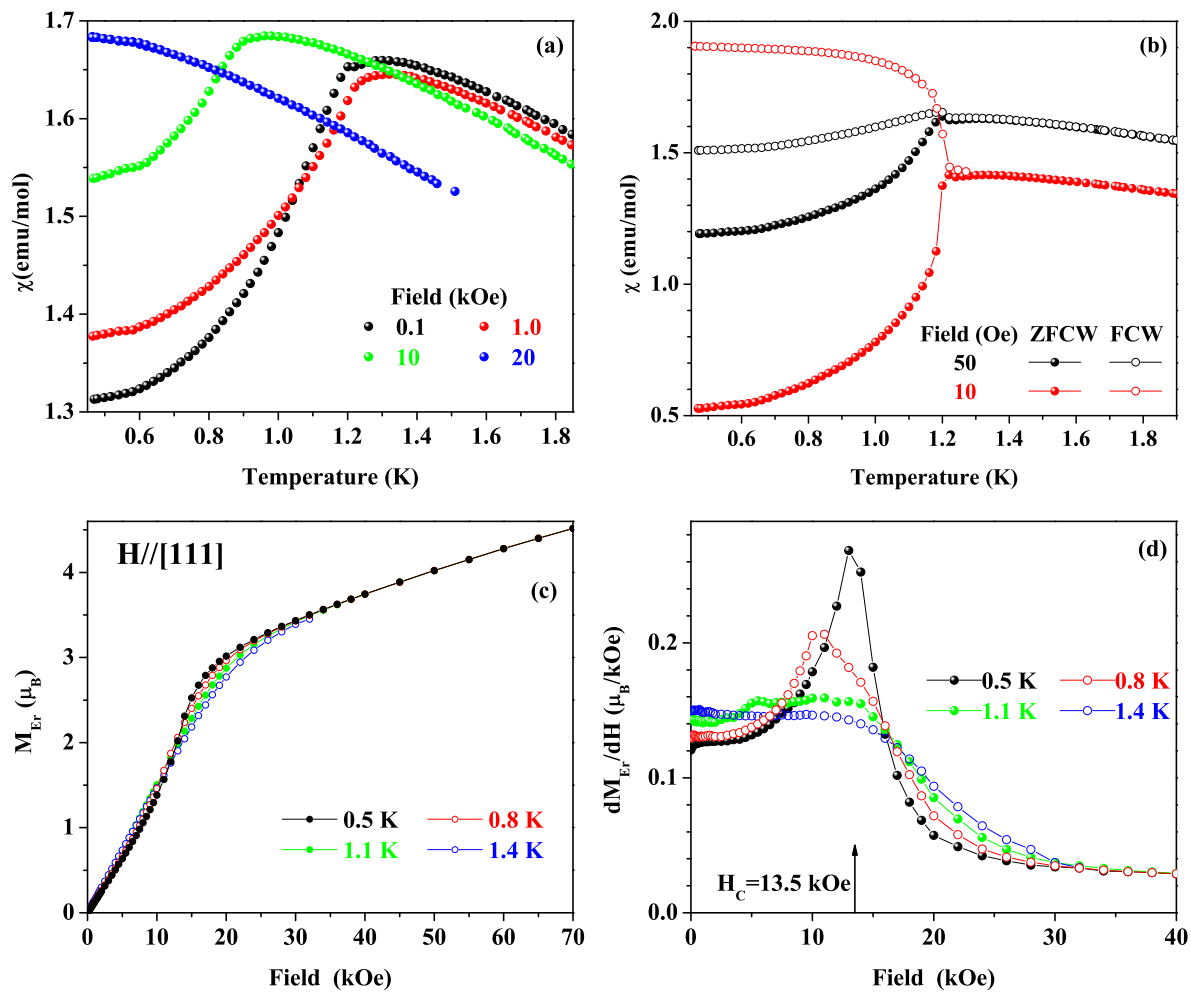}
\caption{Temperature (upper panels) and field (lower panels) dependence of the \mH\ of single crystal of \ETO\ for a field applied along the $[111]$ direction.
		The $\chi(T)$ curves shown in panel (a) have been measured on warming for zero field cooled samples.
		In panel (b) the results for FC and ZFC samples obtained in weak applied fields are contrasted.
		Panel (d) shows the $dM/dH$ curves around a critical field which is temperature dependent and amounts to $H_c=13.5$~kOe at 0.5~K. 
		}
\label{Fig4} 
\end{figure}
In \ETO, the ordering temperature of $T_N=1.2$~K is marked by a change of slope in the temperature dependence of the magnetic susceptibility (see Fig~\ref{Fig4}).
The anomaly moves to lower temperatures and becomes less pronounced in higher fields, disappearing completely above 15~kOe in agreement with the heat capacity measurements~\cite{Ruff_PRL_2008,Sosin_PRB_2010}.
For applied fields between 100~Oe and 20~kOe, no appreciable history dependence has been seen.
For the weaker applied fields, such as 10~Oe and 50~Oe, however, the FC and ZFC susceptibility curves differ significantly (see Fig.~\ref{Fig4}b).
This difference is likely to be associated with the movement of magnetic domain walls in the sample.

\MH\ curves measured in \ETO\ at different temperatures are shown in Fig.~\ref{Fig4}c alongside their derivatives in Fig.~\ref{Fig4}d.
After the initial, nearly linear growth in lower fields, the \mH\ starts to increase much more rapidly around a critical field $H_C$, reaching a value of 3$\mu_B$ per Er ion at 20~kOe.
Then \mH\ continues to increase but at an appreciably lower rate of 0.03$\mu_B$/kOe per Er ion. 
The high field part of the \mH\ appears to be temperature independent, while the influence of temperature is evident around a transition field $H_C$, which is itself temperature dependent and amounts to 13.5~kOe at $T=0.5$~K if defined as the maximum in the $dM/dH(T)$ curves.
The value of the critical field is in perfect agreement with the heat capacity results~\cite{Sosin_PRB_2010}.
Both the diverging behaviour seen in the $dM/dH(T)$ curves at $H_C$ and a considerable increase in the \mH\ above $H_C$ are the further proof that at this field a critical phase transition takes place, most likely quantum in nature~\cite{Ruff_PRL_2008}, rather than a trivial \mH\ saturation.

Although the magnetic properties of \ETO\ are highly anisotropic~\cite{Sosin_PRB_2010}, the \mH\ curves for a field applied along $[110]$ and $[100]$ directions (not shown in this article) appear to demonstrate a field dependence that is functionally  similar to the $[111]$ direction, with a diverging susceptibility around a critical field.
This result is difficult to reconcile with the recent neutron diffraction measurements for a field applied along $[110]$~\cite{Cao_PRB_2010}, where it is claimed that the Er moments have different magnitudes and are aligned along the field at $H_C$, with their values reaching a minimum at this field.
An alternative explanation is that this second-order transition takes place at a field above which the \mH\ process is accompanied by a canting of the magnetic moments off their local ``easy-planes''~\cite{Sosin_PRB_2010}.
It remains to be seen whether an appropriate description of the \mH\ process in \ETO\ can be achieved in a framework of a classical spin system (perhaps by adding a strong easy-plane anisotropy to the exchange Hamiltonian~\cite{Champion_PRB_2003,Glazkov_PRB_2005}) or if a full quantum-mechanical treatment is required.
Despite the large overall magnetic moment on the Er ions,  a quantum influence cannot be ruled out, as the single-ion crystal-field ground state of  Er$^{3+}$ in \ETO\ is a Kramers doublet~\cite{Dasgupta_SSC_2006} with strongly different magnetic moments perpendicular and parallel to the local $\langle 111 \rangle$ axis, which could be effectively described as an $S=1/2$ system.

\subsection{Heisenberg antiferromagnet \GTO}
\begin{figure}
\includegraphics[width=\columnwidth]{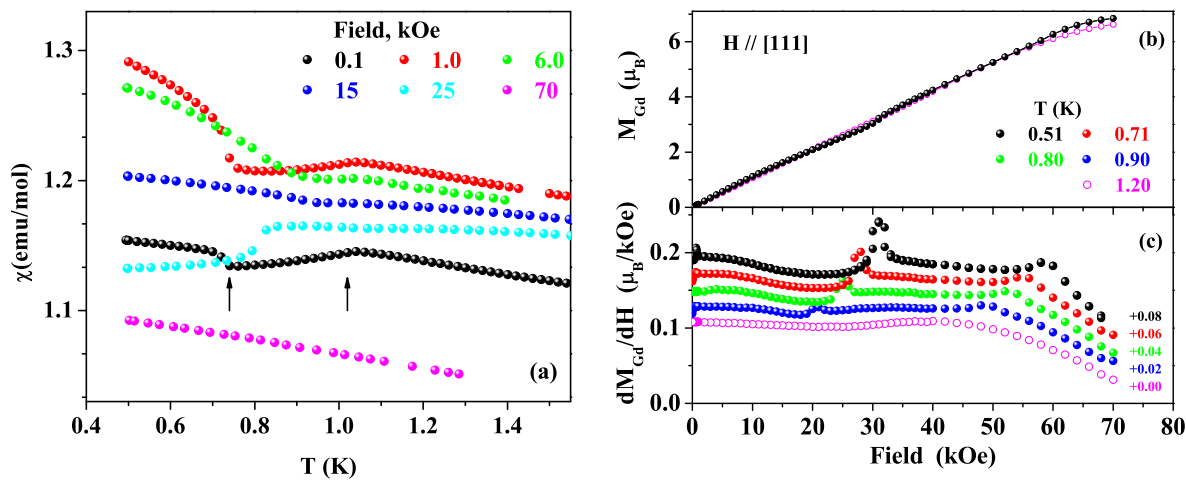}
\caption{Temperature (left) and field (right) dependencies of the \mH\ of \GTO\ single crystal for $H\parallel [111]$.
		The arrows indicate two transition temperatures, $T_{N1}=1.02$~K and $T_{N2}=0.74$~K reported from the specific heat measurements in zero field \cite{Petrenko_PRB_2004}.
		The $dM/dH$ curves obtained from the magnetisation curves at different temperatures have been offset by the specified values for clarity.}
\label{Fig5} 
\end{figure}

Figure~\ref{Fig5}a shows the temperature dependence of the magnetic susceptibility of \GTO\ measured for different fields applied along the $[111]$ axis.
The compound orders magnetically at $T_{N1}=1.02$~K and undergoes a further transition at $T_{N2}=0.74$~K~\cite{Ramirez_PRL_2002,Petrenko_PRB_2004}.
Both these transitions are clearly visible in the $\chi(T)$ curves in low applied field, contrary to the previously published results~\cite{Bonville_JPCM_2003}.
Remarkably, the susceptibility increases with increasing temperature for  $T_{N2}<T<T_{N1}$, while the tendency is opposite for $T>T_{N1}$ and $T<T_{N2}$.
In intermediate magnetic fields either one or two anomalies can be seen in the $\chi(T)$ curves depending on which part of the $H-T$ phase diagram has been traversed~\cite{Petrenko_PRB_2004}, while for magnetic fields above 60~kOe no transition is visible.

Figure~\ref{Fig5}b shows the field dependence of the \mH\ of \GTO\ measured for $H\parallel [111] $ above and below the ordering temperature.
In the highest available field of 70~kOe, at which point the \mH\ is still growing at a considerable rate, a magnetic moment of 6.8$\mu_B$ per Gd ion is reached, close to a maximum value of 7$\mu_B$ expected for a state with $S=7/2$ and $L=0$. 
If a saturation field is defined as the maximum in the $dM/dH(T)$ curve (see Fig.~\ref{Fig5}c), then its value is 59~kOe at $T=0.5$~K.
Apart from the saturation field, $H_{ST}$, another field-induced transition is evident at a field of $H_{ST}/2$, in agreement with the previously reported heat-capacity measurements~\cite{Petrenko_PRB_2004}.
The nature of this transition remains uncertain, as neither of the two proposed models (a canting of the magnetic moments away from their local ``easy-planes''~\cite{Glazkov_JPCM_2007} or a collinear 3 up 1 down structure~\cite{Zhitomirsky_PRL_2000}) seem to fit all the experimental results.
Given that the zero field magnetic structure of \GTO\ is still a matter for debate~\cite{Brammal_arXIv_2010}, it is not surprising that the a theoretical description of the \mH\ process in this compound is still unavailable.

\section{Summary}
In conclusion, we report \mH\ measurements performed at temperatures down to 0.5~K for the highly frustrated \pyro\ systems \HTO, \DTO, \ETO\ and \GTO\ for a field applied along the $[111]$ cubic axis.
In each case unusual phenomena associated with field-induced phase transformations have been observed and commented upon.
The use of high quality single crystal samples allowed for much more accurate and reliable data collection.

\ack The magnetometer used in this research was obtained, through the Science City Advanced Materials project: Creating and Characterising Next Generation Advanced Materials project, with support from Advantage West Midlands (AWM) and part funded by the European Regional Development Fund (ERDF).
The authors also acknowledge financial support from the EPSRC, United Kingdom.


\begin{thebibliography}{99}
\bibitem{demag} Although irregular sample shapes prevent accurate calculations of demagnetisation fields, our estimates show that the difference between applied and effective fields does not exceed 3-4\%.
\end{thebibliography}
\end{document}